\documentclass[sigconf]{acmart}

\newcommand{\added}[1]{{\color{black} #1}}

\newcommand{\motivation}[1]{{\emph{\textbf{M#1}}}}
\newcommand{\situation}[1]{{\emph{\textbf{S#1}}}}
\newcommand{\concern}[1]{{\emph{\textbf{C#1}}}}
\newcommand{\qcode}[1]{{\emph{\textbf{#1}}}}

\newcommand{\tool}{{AI programming assistant\xspace}}

\definecolor{green3}{RGB}{66,179,130}
\definecolor{green2}{RGB}{121,205,169}
\definecolor{green1}{RGB}{196,233,217}

\definecolor{red1}{RGB}{251,219,220}
\definecolor{red2}{RGB}{244,164,166}
\definecolor{red3}{RGB}{236,91,96}

\definecolor{blue1}{RGB}{101,173,246}
\definecolor{blue2}{RGB}{12,112,212}

\definecolor{orange1}{RGB}{250,181,97}
\definecolor{orange2}{RGB}{245,138,7}

\definecolor{purple1}{RGB}{195,176,232}
\definecolor{purple2}{RGB}{146,112,212}

\definecolor{pink1}{RGB}{236,152,223}
\definecolor{pink2}{RGB}{227,100,208}

\definecolor{gray1}{RGB}{220,220,220}

\def\frequencybarchart#1#2#3#4#5#6#7#8{
\resizebox{0.08\linewidth}{7.5pt} {
\begin{tikzpicture}[]
\node[] { \huge \emph{#7}};
\end{tikzpicture}
}
\resizebox {0.81\linewidth} {6.5pt} {%
\begin{tikzpicture}[]
\begin{axis}[
      axis background/.style={fill=gray!30, draw=gray!30},
      axis line style={draw=none},
      tick style={draw=none},
      ytick=\empty,
      xtick=\empty,
      ymin=0, ymax=0.70,
      xmin=0, xmax=6]
\addplot [
      ybar interval=.5,
      fill=green3,
      draw=none,
]
	coordinates {(6*#1,1) (0,0.30)}; %
\addplot [
      ybar interval=.5,
      fill=green2,
      draw=none,
]
	coordinates {(6*(#1+#2),1) (6*#1,1)}; %
\addplot [
      ybar interval=.5,
      fill=gray1,
      draw=none,
]
	coordinates {(6*(#3+#2+#1),1) (6*(#2+#1),1)}; %
\addplot [
      ybar interval=.5,
      fill=red2,
      draw=none,
]
	coordinates {(6*(#4+#3+#2+#1),1) (6*(#3+#2+#1),1)}; %
\addplot [
      ybar interval=.5,
      fill=red3,
      draw=none,
]
	coordinates {(6*(#5+#4+#3+#2+#1),1) (6*(#4+#3+#2+#1),1)}; %
\addplot [
      ybar interval=.5,
      fill=red1,
      draw=none,
]
	coordinates {(6*(#6+#5+#4+#3+#2+#1),1) (6*(#5+#4+#3+#2+#1),1)}; %
\end{axis}%
\end{tikzpicture}%
}
\resizebox{0.08\linewidth}{7.5pt} {
\begin{tikzpicture}[]
\node[] { \huge \emph{#8}};
\end{tikzpicture}
}
}

\def\importancebarchart#1#2#3#4#5#6#7#8{
\resizebox{0.08\linewidth}{7.5pt} {
\begin{tikzpicture}[]
\node[] { \huge \emph{#7}};
\end{tikzpicture}
}
\resizebox {0.81\linewidth} {6.5pt} {%
\begin{tikzpicture}[]
\begin{axis}[
      axis background/.style={fill=gray!30, draw=gray!30},
      axis line style={draw=none},
      tick style={draw=none},
      ytick=\empty,
      xtick=\empty,
      ymin=0, ymax=0.70,
      xmin=0, xmax=6]
\addplot [
      ybar interval=.5,
      fill=blue2,
      draw=none,
]
	coordinates {(6*#1,1) (0,0.30)}; %
\addplot [
      ybar interval=.5,
      fill=blue1,
      draw=none,
]
	coordinates {(6*(#1+#2),1) (6*#1,1)}; %
\addplot [
      ybar interval=.5,
      fill=gray1,
      draw=none,
]
	coordinates {(6*(#3+#2+#1),1) (6*(#2+#1),1)}; %

\addplot [
      ybar interval=.5,
      fill=orange1,
      draw=none,
]
	coordinates {(6*(#4+#3+#2+#1),1) (6*(#3+#2+#1),1)}; %
\addplot [
      ybar interval=.5,
      fill=orange2,
      draw=none,
]
	coordinates {(6*(#5+#4+#3+#2+#1),1) (6*(#4+#3+#2+#1),1)}; %
\addplot [
      ybar interval=.5,
      fill=orange2,
      draw=none,
]
	coordinates {(6*(#6+#5+#4+#3+#2+#1),1) (6*(#5+#4+#3+#2+#1),1)}; %
\end{axis}%
\end{tikzpicture}%
}
\resizebox{0.08\linewidth}{7.5pt} {
\begin{tikzpicture}[]
\node[] { \huge \emph{#8}};
\end{tikzpicture}
}
}

\def\mylegend#1#2{
\resizebox {0.02\linewidth} {6.5pt} {%
\begin{tikzpicture}[]
\begin{axis}[
      axis background/.style={fill=white!30, draw=white!30},
      axis line style={draw=none},
      tick style={draw=none},
      ytick=\empty,
      xtick=\empty,
      ymin=0, ymax=0.70,
      xmin=0, xmax=6]
\addplot [
      ybar interval=.5,
      fill=#2,
      draw=none,
]
	coordinates {(4.5,1) (0,0.30)}; %
\end{axis}%
\end{tikzpicture}%
}%
#1
}

\definecolor{boxcolor}{RGB}{238, 223, 204} %
\DeclareRobustCommand{\mybox}[2][gray!20]{%
\begin{tcolorbox}[   %
        breakable,
        left=0pt,
        right=0pt,
        top=0pt,
        bottom=0pt,
        colback=#1,
        colframe=black,
        width=\dimexpr\columnwidth\relax, 
        enlarge left by=0mm,
        boxsep=5pt,
        outer arc=4pt,
        boxrule=.5mm
        ]
        #2
\end{tcolorbox}
}

\newcommand{\participantQuote}[2]{{
    \parbox{0.95\linewidth}{
        \vspace{2pt}
        \faQuoteLeft\xspace 
        \emph{#1}" (P#2)
    }
}}

\newcommand{\icon}[1]{{\includegraphics[height=1.5\fontcharht\font`\B]{#1}}\xspace}
\newcommand{\meiicon}{\icon{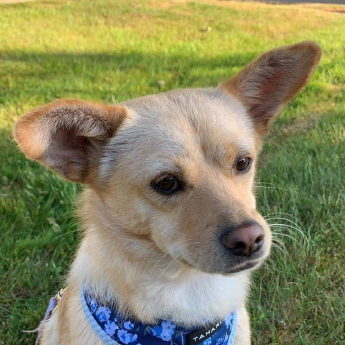}}
\usepackage{xcolor}
\usepackage{xspace}
\usepackage[most]{tcolorbox}
\usepackage{fontawesome}
\usepackage{pgfplots}
\usepackage{pgfplotstable}
\usepackage{multirow}
\usepackage{multicol}

\AtBeginDocument{%
  }

\copyrightyear{2024} 
\acmYear{2024} 
\setcopyright{rightsretained} 
\acmConference[ICSE 2024]{2024 IEEE/ACM 46th International Conference on Software Engineering}{April 14--20, 2024}{Lisbon, Portugal}
\acmBooktitle{2024 IEEE/ACM 46th International Conference on Software Engineering (ICSE 2024), April 14--20, 2024, Lisbon, Portugal}\acmDOI{10.1145/3597503.3608128}
\acmISBN{979-8-4007-0217-4/24/04}

\begin{document}

\title{A Large-Scale Survey on the Usability of AI Programming Assistants: Successes and Challenges}

\author{Jenny T. Liang}
\affiliation{%
  \institution{Carnegie Mellon University}
  \city{Pittsburgh, PA}
  \country{USA}}
\email{jtliang@cs.cmu.edu}

\author{Chenyang Yang}
\affiliation{%
  \institution{Carnegie Mellon University}
  \city{Pittsburgh, PA}
  \country{USA}}
\email{cyang3@cs.cmu.edu}

\author{Brad A. Myers}
\affiliation{%
  \institution{Carnegie Mellon University}
  \city{Pittsburgh, PA}
  \country{USA}}
\email{bam@cs.cmu.edu}

\renewcommand{\shortauthors}{Liang et al.}

\begin{abstract}
The software engineering community recently has witnessed widespread deployment of \tool{}s, such as GitHub Copilot. 
However, in practice, developers do not accept AI programming assistants' initial suggestions at a high frequency. 
This leaves a number of open questions related to the usability of these tools.
To understand developers' practices while using these tools and the important usability challenges they face, we administered a survey to a large population of developers and received responses from a diverse set of 410 developers.
Through a mix of qualitative and quantitative analyses, we found that developers are most motivated to use \tool{}s because they help developers reduce key-strokes, finish programming tasks quickly, and recall syntax, but resonate less with using them to help brainstorm potential solutions.
We also found the most important reasons why developers do \textit{not} use these tools are because these tools do not output code that addresses certain functional or non-functional requirements and because developers have trouble controlling the tool to generate the desired output.
Our findings have implications for both creators and users of \tool{}s, such as designing minimal cognitive effort interactions with these tools to reduce distractions for users while they are programming.

\end{abstract}



\maketitle

\section{Introduction}
The recent widespread deployment of \tool{}s, such as GitHub Copilot~\cite{github2023copilot} and ChatGPT~\cite{openai2023chatgpt}, has introduced a new paradigm to building software that has taken the software engineering community by storm. Some current publications report that \tool{}s are powerful enough to produce high-quality code suggestions for developers~\cite{xu2022systematic, yetistiren2022assessing}.
While some recent studies do not find any significant difference in using \tool{}s in terms of task completion~\cite{vaithilingam2022expectation,xu2022ide} and code quality~\cite{imai2022github}, other studies find these tools are positively associated with developers' self-perceived productivity~\cite{ziegler2022productivity}.

\begin{figure}[t!]
\centering
\includegraphics[trim=0 225 55 0, clip, width=\linewidth, keepaspectratio]{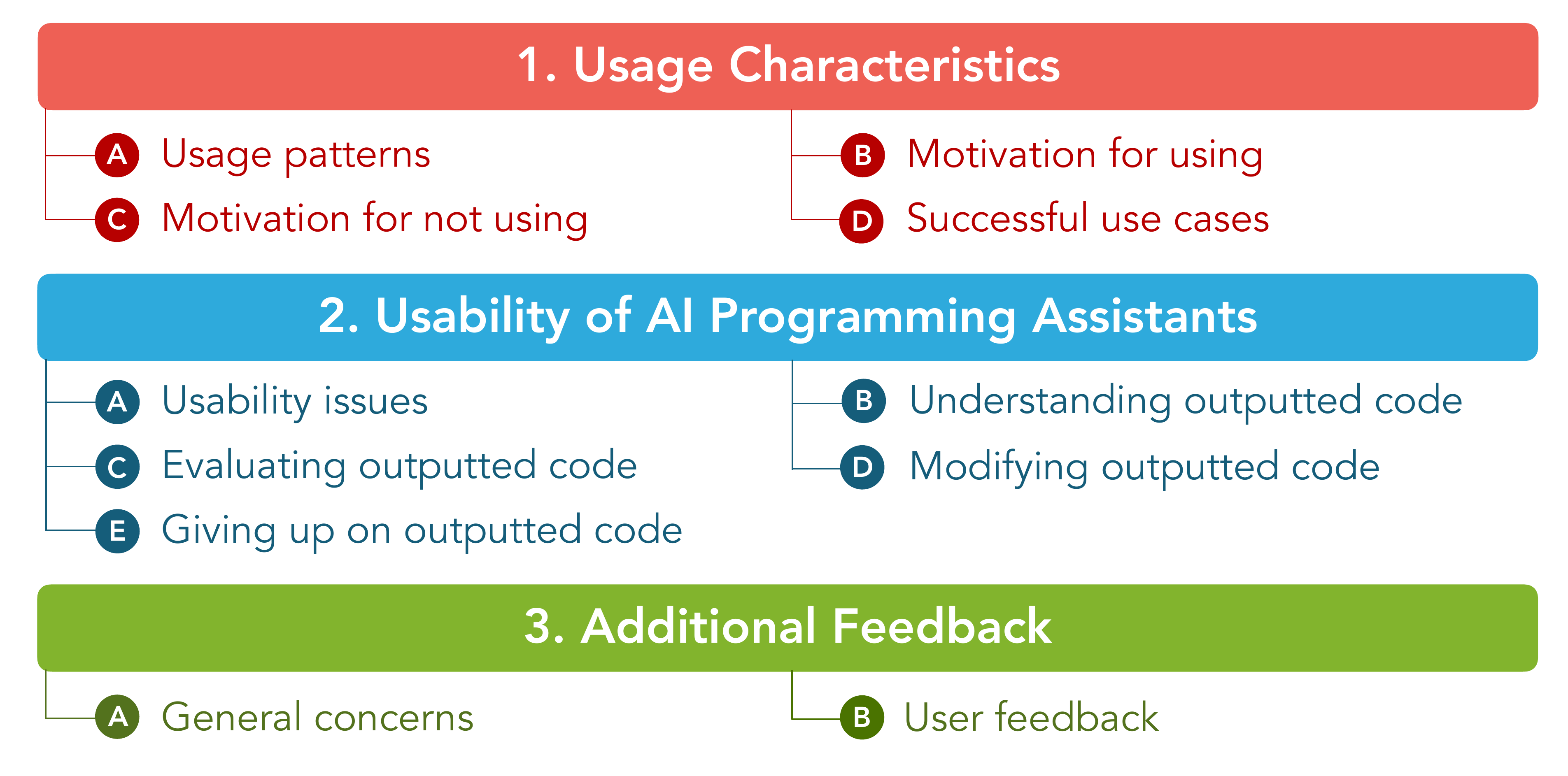} 
\caption{An overview of the topics covered in our usability study of \tool{}s.}
\label{fig:overview}
\end{figure}

However, in practice, prior literature indicates that developers do not accept \tool{}s' initial suggestions at a high frequency. 
~\citet{ziegler2022productivity} found that developers accepted 23.3\%, 27.9\%, and 28.8\% of GitHub Copilot's suggestions for TypeScript, JavaScript, and Python respectively. 
There are many potential reasons for the lack of adoption of \tool{}s' suggestions.
One study shows that developers feel concerned that the generated code may contain defects, may not adhere to the project's coding style, or may be difficult to understand~\cite{vaithilingam2022expectation}. 
Other studies report that software developers face barriers in comprehending and debugging generated code to fit their use cases, because they need to have prior knowledge of the underlying programming principles, frameworks, or APIs~\cite{xu2022ide,barke2022grounded}.

While prior work has surfaced initial results about the usability of state-of-the-art \tool{}s, to our knowledge, they have not systematically investigated the prevalence of usability factors related to these tools. 
Quantifying the usability of \tool{}s could help tool creators understand which usability aspects are currently successful in practice. Further, it could help tool creators prioritize features and improvements to the modeling and user interface of these tools in the future, potentially increasing the adoption of these tools and improving the productivity of developers. 
Usability is an important factor to study in \tool{}s, since modeling improvements may not necessarily address the needs of developers, rendering these tools hard-to-use or even useless~\cite{myers2016programmers}.

We performed an exploratory qualitative study in January 2023 to understand developers' practices when using \tool{}s and the importance of the usability challenges that they face. 
We used a survey as a research instrument to collect large-scale data on these phenomena to understand their importance to the usability of \tool{}s (see Figure~\ref{fig:overview}).

In the end, we collected and analyzed responses from 410 developers who were recruited from GitHub repositories related to \tool{}s, such as GitHub Copilot and Tabnine~\cite{tabnine2023tabnine}. 
In summary, we find that: 

\vspace{1em}

\noindent
\textbf{Usage characteristics of \tool{}s} (Section~\ref{sec:usage-characteristics})
\begin{enumerate}
        \item Developers who use GitHub Copilot report a median of 30.5\% of their code being written with help from the tool.
        \item Developers report the most important reasons why they use \tool{}s are because of the tools' ability to help developers reduce key-strokes, finish programming tasks quickly, and recall syntax.
        \item The most important reasons why developers do \textit{not} use these tools at all are that the tools generate code that do not meet certain functional or non-functional requirements and that it is difficult to control these tools to generate the desired output.
    \end{enumerate}

\noindent
\textbf{Usability of \tool{}s} (Section~\ref{sec:usability-challenges})
    \begin{enumerate}
        \addtocounter{enumi}{3}
        \item Developers report the most prominent usability issues are that they have trouble understanding what inputs cause the tool's generated code, giving up on incorporating the outputted code, and controlling the tool to generate helpful code suggestions.
        \item The most frequent reasons why users of these tools give up on using outputted code are that the code does not perform the correct action or it does not meet functional or non-functional requirements. 
    \end{enumerate}

\noindent
\textbf{Additional feedback about \tool{}s from users} (Section~\ref{sec:user-feedback})
    \begin{enumerate}
        \addtocounter{enumi}{5}
        \item Developers would like to improve their experience with \tool{}s by providing feedback to the tool to correct or personalize the model as well as by having these tools to learn a better understanding of code context, APIs, and programming languages.
    \end{enumerate}

In this paper, we refer to \emph{tool creators} as the individuals who build and develop software related to \tool{}s. 
\emph{Tool users} are the people who use these tools while building software. We use this term interchangeably with \emph{developers}. 
Finally, we use the term \emph{inputs} to refer to the code and natural language context \tool{}s use to produce \emph{outputted code}, which we also call \emph{generations}.

\section{Related Work}
\label{sec:related-work}

We discuss work related to the usability of \tool{}s. Since this field is rapidly developing, the papers discussed are a snapshot of the current progress in the field as of March 2023. 

Prior work includes a few usability studies on various \tool{}s using programming by demonstration approaches~\cite{ferdowsifard2020small,chasins2018rousillon} and recurrent neural networks-based approaches \cite{lin2017program}. 
\citet{lin2017program} reported that developers have difficulty in correcting generated code, while ~\citet{ferdowsifard2020small} showed that a mismatch in the perceived versus actual capabilities of program synthesizers may prevent the user from using them effectively. 
Meanwhile, ~\citet{jayagopal2022exploring} also conducted usability studies to understand the learnability of five of these tools with novices. 
Finally, ~\citet{mcnutt2023design} enumerated a design space of interactions with code assistants, including how users can disambiguate programs or refine generated code. 
Our study diverges from these works by evaluating \tool{}s that are widely used in practice by developers rather than evaluating these tools in laboratory settings. 
In particular, we examine tools based on the transformer neural network architecture~\cite{vaswani2017attention}, such as GitHub Copilot and Tabnine. Transformer-based tools have shown strong performance in working with both natural language and code inputs~\cite{xu2022systematic} compared to other types of these tools.

Researchers have performed user studies on transformer-based \tool{}s~\cite[e.g.,][]{xu2022ide, jiang2022discovering}.
Both studies found users may have trouble expressing the intent in their queries. 
In particular, \citet{xu2022ide} revealed a challenge their users faced was that the tool assumed background knowledge in underlying modules or frameworks. 

Also related to our study are usability studies on how users are using GitHub Copilot in practice. ~\citet{vaithilingam2022expectation} performed a user study of GitHub Copilot with 24 participants, where they found users struggled with understanding and debugging the generated code. 
In a user study with 20 participants, ~\citet{barke2022grounded} found that developers used GitHub Copilot in two different modes--when they do not know what to do and explore different options (i.e., \emph{exploration mode}), or when they do know what to do but use GitHub Copilot to complete the task faster (i.e., \emph{acceleration mode})--and that users are less willing to modify suggestions. 
Meanwhile, ~\citet{mozannar2022reading} identified 12 core activities associated with using GitHub Copilot, such as verifying suggestions, looking up documentation, and debugging code, which was then validated on a user study with 21 developers. 
Finally, ~\citet{ziegler2022productivity} performed a large-scale user study of GitHub Copilot. They analyzed telemetry data from the model and 2,631 survey responses on developers' perceived productivity with the tool. They reported that 23.3\%, 27.9\%, and 28.8\% of GitHub Copilot's suggestions were accepted for TypeScript, JavaScript, and Python respectively, and 22.2\% for all other languages. 
We extend their user study by performing a large scale study with a focus on the usability challenges of many \tool{}s, including GitHub Copilot, which provides possible explanations for their findings.

Other works have studied various design aspects of \tool{}s. For instance, ~\citet{vaithilingam2023towards} suggested six design principles of inline code suggestions from \tool{}s, such as having glanceable suggestions.  
With the recent popularity of transformer-based chatbots, such as ChatGPT~\cite{openai2023chatgpt}, recent work ~\cite[e.g.,][]{robe2022pair, ross2023programmer} has investigated developers' interactions with conversational chatbots. 
For example, ~\citet{ross2023programmer} find that developers are initially skeptical of chatbot programming assistants, but are hopeful about their ability to improve their productivity after using them.

Many of the user studies enumerate potential usability challenges of using \tool{}s. However, it is unclear to what extent the enumerated challenges are important to developers in practice. Therefore, our study validates and extends these works by quantifying to what extent these usability challenges are encountered by developers in practice. Compared to prior work, we also investigate a larger number of these tools and have a broader focus on usability of both the tools and the tool's outputted code.

\section{Methodology}
\label{sec:methodology}

\subsection{Participants}
\label{sec:participants}
We recruited a large number of participants in order to elicit a diverse range of programming experiences. 

\subsubsection*{Sampling strategy}
\label{sec:sampling-strategy}
We recruited participants by selecting contributors from GitHub repositories, following a sampling strategy similar to prior work~\cite{liang2022understanding,huang2021leaving}. To recruit developers who are interested in \tool{}s, we identified the three projects related to these tools. Two were from GitHub's official GitHub account (i.e., \emph{github/copilot-docs}~\cite{github2023copilotdocs} and \emph{github/copilot.vim}~\cite{github2023copilotvim}), while one was the official project repository for Tabnine~\cite{tabnine2023tabnine}, a popular \tool\xspace (i.e., \emph{codota/Tabnine}~\cite{github2023tabnine}).
To sample participants from the repositories, we used GitHub's GraphQL API~\cite{github2023graphql} to retrieve users who had forked or starred the repositories.
2,329 GitHub users forked, 21,302 GitHub users starred, and 396 GitHub users watched \emph{github/copilot-docs}. 379 GitHub users forked, 6,299 GitHub users starred, and 87 GitHub users watched \emph{github/copilot.vim}. 420 GitHub users forked, 9,594 GitHub users starred, and 133 GitHub users watched \emph{codota/Tabnine}.
We then took the set union of the 9 sets of participants, removing all duplicates.
This resulted in 33,983 unique GitHub users who had activities associated with the three repositories.

Finally, we filtered the GitHub users by whether they had a publicly available email address, yielding 10,530 unique users who we invited to take the survey. A random sample of 500 users was first sent the survey to verify the quality of the data. Email invitations were sent to the remaining 10,030 users.

\subsubsection*{Demographics}
The Qualtrics survey was sent to all 10,530 GitHub users and received 410 responses, resulting in a response rate of around 4\%. This response rate is comparable to other research surveys in software engineering~\cite[e.g.,][]{liang2022understanding, smith2013improving}. \

We summarize the attributes of our participants. Questions on their background were optional and thus may not sum up to 410. Overall, participants represented 57 unique countries. They were from Africa ($n = 9$), Asia ($n = 116$), Europe ($n = 77$), North America ($n = 77$), Oceania ($n = 4$), and South America ($n = 13$). They also represented multiple genders, such as man ($n = 280$), woman ($n = 8$), and non-binary ($n = 7$). Participants programmed under a variety of contexts, including for their profession as a software engineer ($n = 203$) or an end-user developer ($n = 82$), an open-source project ($n = 131$), hobby ($n = 155$), and/or school ($n = 172$). Additionally, they had a wide range of programming experience, ranging from 1 to 41 years, with a median of 6 years. Survey participants reported using a variety of programming languages, such as Python ($n = 199$), JavaScript ($n = 175$), HTML/CSS ($n = 157$), TypeScript ($n = 123$), Bash/Shell ($n = 134$), and/or Java ($n = 84$). They also used \tool{}s (see Table~\ref{tab:usage}), such as GitHub Copilot, Tabnine, Amazon CodeWhisperer, ChatGPT, and \tool{}s specific to an organization that was trained on proprietary code.

\begin{figure}
    \begin{tcolorbox}[left=-8pt,right=2pt,top=2pt,bottom=2pt]
    {\sc \hspace{10pt}Survey Questions}
    \begin{itemize}
        \item For this software project, estimate what percent of your code is written with the help of the following code generation tools.
        \item For each of the following reasons why you use code generation tools in this software project, rank its importance.
        \item For each of the following reasons why you do not use code generation tools, rank its importance.
        \item For your software project, estimate how often you experience the following scenarios when using code generation tools.
        \item For your software project, estimate how often the following reasons are why you find yourself giving up on code created by code generation tools.
        \item[{\tiny\faStar}] What types of feedback would you like to give to code generation tools to make its suggestions better? Why?
    \end{itemize}
    \end{tcolorbox}
    \vspace{-0.5\baselineskip}
    \caption{A subset of the actual survey questions about the usability of \tool{}s. An open-ended question is indicated with a star ({\tiny\faStar}). The complete survey instrument is in the supplemental materials~\cite{supplemental-materials}.}
    \label{fig:survey}
\end{figure}

\subsection{Survey}
\label{sec:survey}
We designed a 15-minute Qualtrics survey to gather data for our research questions and distributed it to participants using the sampling strategy described in Section~\ref{sec:sampling-strategy}. After completing the survey, participants could join a sweepstakes to win one of four \$100 electronic gift certificates. All questions in the survey were optional. The study was approved by our institution's institutional review board.

The survey first asked participants how often they used \tool{}s and whether they had any concerns about using these tools. If the participant used \tool{}s, they were asked to consider a specific project where they used \tool{}s and were asked a set of questions regarding their experience with these tools. Survey topics included: why participants use \tool{}s, how often these tools are used, strategies participants use to make \tool{}s work better, and why participants give up using generated code. If the participant did not use \tool{}s, they answered questions regarding why they did not. 

The survey also collected information on the participants' programming backgrounds and demographics. Following best practices, we used the HCI Guidelines for Gender Equity and Inclusivity to collect gender-related information~\cite{scheuerman2020hci}. We allowed participants to select multiple responses for questions on gender. A subset of the survey questions is included in Figure~\ref{fig:survey}; the full survey instrument is included in the supplemental materials~\cite{supplemental-materials}. While developing the survey, an external researcher reviewed and provided feedback on the survey for clarity and topic coverage.

We conducted pilots of the survey to identify and reduce confounding factors, following the best practices for experiments with human subjects in software engineering research~\cite{ko2015practical}. We piloted drafts of the survey with 11 developers, \added{who were recruited through snowball sampling}. These pilots helped clarify wording, ensure data quality, and identify usability factors prior literature may have missed. The survey was updated between each round of feedback. \added{The results from the pilots were not included in the data used in this study.}

\begin{table*}
  \centering

\small

\caption{Participants' self-reported usage of popular \tool{}s. An asterisk (*) denotes a write-in suggestion, which has limited information on its usage distribution. Percentages in italics on the chart ($N$\emph{\%}) represent the percent of the distribution that reported "Always"/"Often" (left) and "Rarely"/"Tried but gave up" (right).}
\label{tab:usage}
\begin{tabular}{p{0.32\linewidth}p{0.05\linewidth}p{0.09\linewidth}|p{0.45\linewidth}}
\toprule
\multirow{2}{*}{\textbf{Tool}} & \multirow{2}{*}{\textbf{\# users}} & \hfil \textbf{Med. \%} & \multirow{2}{*}{\textbf{Usage distribution}} \\
& & \textbf{code written} & \\
\midrule
Amazon CodeWhisperer & \hfil 50 & \hfil 5\% & \frequencybarchart{0.15}{0.09}{0.15}{0.23}{0.38}{0}{24\%}{61\%}\\
ChatGPT* & \hfil 25 & \hfil 20\% & \frequencybarchart{0.45}{0.14}{0.27}{0.14}{0}{0}{59\%}{14\%} \\
GitHub Copilot & \hfil 306 & \hfil 30.5\% & \frequencybarchart{0.43}{0.13}{0.14}{0.08}{0.22}{0}{46\%}{30\%} \\
TabNine & \hfil 118 & \hfil 20\% & \frequencybarchart{0.16}{0.11}{0.08}{0.14}{0.52}{0}{27\%}{66\%} \\
\hangindent=1em Organization-specific code generation tool trained on proprietary code & \hfil 54 & \hfil 37\% & \frequencybarchart{0.18}{0.11}{0.15}{0.3}{0.26}{0}{29\%}{56\%} \\
\midrule
\multicolumn{4}{c}{\mylegend{Always (1+ times daily)}{green3} \mylegend{Often (once daily)}{green2} \mylegend{Sometimes (weekly)}{gray1}\mylegend{Rarely (monthly)}{red2} \mylegend{Tried but gave up}{red3}} \\
\bottomrule
\end{tabular}
\end{table*}

\subsection{Analysis}
\label{sec:analysis}
To analyze the data, we used both quantitative and qualitative techniques. This is because survey questions were largely closed-ended but participants could also select an "other" option, and many questions also provided space to enter open-ended responses. 
The choices are based on survey piloting and results from prior literature on human evaluations of \tool{}s (i.e., ~\cite{barke2022grounded, imai2022github, puryear2022github, denny2023conversing, vaithilingam2022expectation, cheng2022would, ziegler2022productivity, dakhel2022github, jayagopal2022exploring, jiang2022discovering, xu2022ide}). \added{The first author reviewed these papers and extracted mentions of usability-related issues with \tool{}s, resulting in a set of usability issues with these tools. This set of usability issues was then de-duplicated and used as choices for closed-ended questions in the survey.}
Below, we describe our methods in further detail.

\subsubsection*{Quantitative analysis} 
To perform quantitative analysis on the closed-ended questions, we followed best practices for statistical analysis techniques described by Kitchenham and Pfleeger on how to analyze survey data~\cite{survey-guidelines}. In particular, we report the frequencies of how often an item was selected. We also report how frequently participants rated statements as being important or very important, situations as occurring often or always, and feeling concerned or very concerned about a situation. Following best practices~\cite{myers2016programmers}, we report measurements on perceived frequency to understand the importance of a situation rather than an accurate measurement on how frequently a situation occurs.

\subsubsection*{Qualitative analysis}

For qualitative analysis, the first two authors performed multiple rounds of open coding on each set of responses to the open-ended questions.
We used general best practices~\cite{hammer2014confusing,saldana2009coding}, such as interpreting generated codes as itemized claims about the data to be investigated in other work and shuffling responses to reduce any ordering effects that could emerge during coding.

In the first round of coding, the authors open-coded the same initial set of 100 responses. 
Each response was labeled with zero or more codes. Each code was given a unique identifier and brief description. 
Then, the authors convened to discuss the resulting set of codes and their scopes.
To merge the codes, the authors identified codes with similar themes and merged them into a single code in the shared codebook. 
The remaining codes were then added to or removed from the codebook by a unanimous vote between the two authors.
Coding disagreements most frequently occurred due to different scopes of codes rather than the meaning of participants' statements.
The authors then jointly performed a second round of coding on the original data by applying codes from the shared codebook onto each instance based on a unanimous vote.
We do not report IRR because following best practices from~\citet{hammer2014confusing}, each instance's codes were unanimously agreed upon and because the codes were the process, not the product~\cite{mcdonald2019reliability}.

\begin{table*}
  \centering

\small

\caption{Participants' motivations for using and not using \tool{}s. }
\label{tab:usage-motivation}
\begin{tabular}{p{0.02\linewidth}p{0.47\linewidth}|p{0.45\linewidth}}
\toprule
& \textbf{Motivation} & \textbf{Distribution} \\
\midrule
& \textbf{\emph{A. For using}} & \\
M1 & To have an autocomplete or reduce the amount of keystrokes I make. & \importancebarchart{0.55}{0.31}{0.08}{0.04}{0.02}{0}{86\%}{6.2\%}\\
M2 & To finish my programming tasks faster. & \importancebarchart{0.43}{0.33}{0.12}{0.1}{0.02}{0}{76\%}{12\%}\\
M3 & \hangindent=1em To skip needing to go online to find specific code snippets, programming syntax, or API calls I'm aware of, but can't remember. & \importancebarchart{0.41}{0.28}{0.17}{0.1}{0.04}{0}{68\%}{14\%}\\
M4 & \hangindent=1em To discover potential ways or starting points to write a solution to a problem I'm facing. & \importancebarchart{0.24}{0.26}{0.26}{0.15}{0.09}{0}{50\%}{24\%}\\
M5 & To find an edge case for my code I haven't considered. & \importancebarchart{0.15}{0.22}{0.19}{0.25}{0.19}{0}{36\%}{44\%}\\
\midrule
& \textbf{\emph{B. For not using}} & \\
M6 & \hangindent=1em Code generation tools write code that doesn't meet functional or non-functional (e.g., security, performance) requirements that I need. & \importancebarchart{0.31}{0.24}{0.12}{0.18}{0.16}{0}{54\%}{34\%} \\
M7 & It's hard to control code generation tools to get code that I want. & \importancebarchart{0.24}{0.24}{0.18}{0.2}{0.16}{0}{48\%}{36\%} \\
M8 & \hangindent=1em I spend too much time debugging or modifying code written by code generation tools. & \importancebarchart{0.18}{0.2}{0.18}{0.16}{0.29}{0}{38\%}{45\%} \\
M9 & I don't think code generation tools provide helpful suggestions. & \importancebarchart{0.14}{0.2}{0.22}{0.22}{0.24}{0}{34\%}{46\%} \\
M10 & I don't want to use a tool that has access to my code. & \importancebarchart{0.24}{0.06}{0.2}{0.1}{0.41}{0}{30\%}{51\%} \\
M11 & \hangindent=1em I write and use proprietary code that code generation tools haven't seen before and don't generate. & \importancebarchart{0.08}{0.2}{0.14}{0.22}{0.37}{0}{28\%}{59\%} \\
M12 & To prevent potential intellectual property infringement. & \importancebarchart{0.13}{0.13}{0.09}{0.1}{0.56}{0}{26\%}{66\%} \\
M13 & I find the tool's suggestions too distracting. & \importancebarchart{0.14}{0.12}{0.24}{0.22}{0.29}{0}{26\%}{51\%} \\
M14 & I don't understand the code written by code generation tools. & \importancebarchart{0.02}{0.14}{0.08}{0.27}{0.49}{0}{16\%}{76\%} \\
M15 & I don't want to use open-source code. & \importancebarchart{0.02}{0.08}{0.08}{0.13}{0.76}{0}{10\%}{89\%} \\
\midrule
\multicolumn{3}{c}{\mylegend{Very important}{blue2} \mylegend{Important}{blue1} \mylegend{Moderately important}{gray1}\mylegend{Slightly important}{orange1} \mylegend{Not important at all}{orange2}} \\
\bottomrule
\end{tabular}
\end{table*}

\section{Usage Characteristics}
\label{sec:usage-characteristics}
We present our findings on how developers use \tool{}s. 
We first present quantitative results on how developers use these tools (Section~\ref{sec:usage-patterns}) and developers' motivations for using them (Section~\ref{sec:motivation}). To elucidate the quantitative results, we describe qualitative results on successful use cases (Section~\ref{sec:successful-use-cases}) and users' strategies to generate helpful output (Section~\ref{sec:user-input-strategies}).

\subsection{Usage patterns} 
\label{sec:usage-patterns}
In the survey, we asked participants to describe how often they used \tool{}s and how much of their code was written with the help of these tools (see Table~\ref{tab:usage}). We report the median percentage of code written by each tool's users. 
Unsurprisingly, GitHub Copilot was the most popular \tool{} by the number of users (306), with 46\% of its users reporting using the tool frequently. 
GitHub Copilot's users reported writing 30.5\% of their code with the help of the tool.
However, organization-specific \tool{}s helped write the largest percentage of code for survey participants (37\%). 
Interestingly, we found that chatbot-based programming assistants (i.e., ChatGPT) were self-reported by 25 participants. 
Even though ChatGPT had the highest proportion of frequent users (59\%), it was the penultimate tool in terms of the amount of code it helped write for survey participants (20\%). 

\subsection{Motivation}
\label{sec:motivation}
\subsubsection*{Motivation for using}
Participants who reported using an \tool{} on at least a monthly basis reported their motivations for using these tools (see Table~\ref{tab:usage-motivation}-A). Participants largely used these tools for convenience in programming--86\%, 76\%, and 68\% of participants cited an important motivation for using these tools was autocompletion (\motivation{1}), finishing tasks faster (\motivation{2}), and skipping going online to recall syntax respectively (\motivation{3}). On the other hand, 50\% and 36\% of participants said an important reason for using these tools was finding potential code solutions (\motivation{4}) or edge cases respectively (\motivation{5}).

\subsubsection*{Motivation for not using}
Participants who reported not using any \tool{} on at least a monthly basis reported their motivations for \emph{not} using these tools (see Table~\ref{tab:usage-motivation}-B). 
Participants seemed to not use these tools because the tools did not provide useful or relevant output.
Two important motivations were that the models did not write code that met certain functional or non-functional requirements (\motivation{6}, 54\%) and users had difficulty controlling the model (\motivation{7}, 48\%). 34\% of participants cited these tools not providing helpful suggestions as an important reason for not using them (\motivation{9}). By having code that was not useful, users engaged in the time-consuming process of modifying or debugging code (\motivation{8}). This was also a salient motivation, as 38\% of participants rated it as an important reason for not using these tools.
Participants resonated the least with not understanding generated code (\motivation{14}) and not wanting to use open-source code (\motivation{15}), as 76\% and 89\% of participants rated them as not important.

\subsection{Successful use cases}
\label{sec:successful-use-cases}
Survey participants described situations where they were most successful in using \tool{}s. We found 10 types of situations, which we describe below. We report the frequencies of the codes using the multiplication symbol ($\times$).

\paragraph{\textbf{Repetitive code} (78$\times$)}
Participants were successful in using the \tool{}s to generate repetitive code, such as \emph{"boilerplate [code]"} (P165), \emph{"repetitive endpoints for crud"} (P164), and \emph{"college assignments"} (P265) that had repeated functionality or were common programming tasks. This was the most frequent code in our data.

\participantQuote{Complete code that is highly repetitive but cannot be copied and pasted directly.}{195}

\paragraph{\textbf{Code with simple logic} (68$\times$)}
Consistent with prior work~\cite{vaithilingam2022expectation}, participants reported using \tool{}s to successfully generate code with simple logic. This was the second most mentioned code in the dataset. 
Examples include \emph{"small independent utils functions"} (P155), \emph{"sorting algorithms"} (P177), and \emph{"small functions like storing the training model into local file systems"} (P255). 
Participants said that having the tool write more complex logic often resulted in it not working:

\participantQuote{It however, fails assisting me when I'm writing a more complex algorithm (if not well known).}{28}

\paragraph{\textbf{Autocomplete} (28$\times$)}
We found participants also utilized \tool{}s to do short autocompletions of code, which is associated most with \emph{acceleration mode} usages of these tools~\cite{barke2022grounded}. 
This code was the third most mentioned code in the dataset.

\participantQuote{I wrote \texttt{s\_1, a\_1 = draw('file\_1')}, then I want to complete \texttt{s\_2, a\_2 = draw('file\_2')}. After I type \texttt{s\_2}, copilot helps me [with] this line.}{240}

\paragraph{\textbf{Quality assurance} (21$\times$)}
Participants reported using \tool{}s for quality assurance, such as \emph{"[generating] useful log messages"} (P212) and \emph{"[producing] a lot of test cases quickly"} (P356). As found in prior work~\cite{barke2022grounded}, participants used these tools to consider edge cases:

\participantQuote{This tool can almost instantly generate the code with good edge case coverage.}{160}

\paragraph{\textbf{Proof-of-concepts} (20$\times$)}
Similar to prior work~\cite{barke2022grounded, vaithilingam2022expectation, xu2022ide}, participants mentioned that using \tool{}s helped with brainstorming or building proof-of-concepts by helping generate multiple implementations for a given problem. Participants relied on this when they \emph{"need[ed] another solution"} (P193) or \emph{"only [had] a fuzzy idea about how to approach it"} (P163), so these tools also helped with provide a starting implementation to work off of: 

\participantQuote{We most use these tools at the beginning as a start point or when we get stuck.}{21}

\paragraph{\textbf{Learning} (19$\times$)}
Study participants also utilized these tools when \emph{"learning new programming languages"} (P197) or \emph{"new libraries"} (P140) they had limited to no experience with, rather than using online documentation~\cite{rao2020analyzing} or video tutorials~\cite{macleod2015code}. Participants reported that it was especially useful when a project used multiple programming languages:

\participantQuote{Since [the codebase] is a polyglot project with golang, java, and cpp implementations, I benefit a lot from...polyglot support.}{40}

\paragraph{\textbf{Recalling} (19$\times$)}
As found in prior work~\cite{xu2022ide}, participants leveraged \tool{}s to find syntax of programming languages or API methods that they were familiar with, but could not recall. This replaced the traditional methods of using web search~\cite{rao2020analyzing} to find online resources like StackOverflow~\cite{herbsleb2001global,mamykina2011design} to recall code snippets or syntax:

\participantQuote{To skip needing to go online to find...code snippets.}{179}

\paragraph{\textbf{Efficiency} (18$\times$)}
Study participants also echoed prior work~\cite{ziegler2022productivity} by describing an \tool's ability to \emph{"speed up...work"} (P246). 
Participants reported that it helped them to "stay in the flow", an important aspect of developer productivity~\cite{forsgren2021space}: 

\participantQuote{Code generation will help the process go smoother and does not introduce unwanted interruptions.}{166}

\paragraph{\textbf{Documentation} (6$\times$)}
A few participants used \tool{}s to generate documentation. One participant noted generating documentation helped with collaboration:

\participantQuote{I mainly use it to...annotate my code for my colleagues.}{258}

\paragraph{\textbf{Code consistency} (4$\times$)}
A few participants used these tools to improve style consistency in a codebase, which is a factor developers consider while making implementation decisions~\cite{liang2023qualitative}. Participants applied these tools to \emph{"[follow]...standard clean code style"} (P156), such as \emph{"proper indentation in different [programming] languages"} (P50). It also helped with consistency within a project:

\participantQuote{To ensure consistency of code by quickly referencing sources created within the project.}{36}

\subsection{User input strategies}
\label{sec:user-input-strategies}
Finally, we asked participants to enumerate strategies they used to get \tool{}s to output the best answers. We found 7 strategies, which we describe below.

\paragraph{\textbf{Clear explanations} (99$\times$)}
The most popular strategy participants reported was providing very clear and explicit explanations of what the code should do in comments, which is a major activity while using \tool{}s~\cite{mozannar2022reading}. Participants wrote \emph{"a docstring which tells the function of the function"} (P22) or \emph{"outlining preconditions and postconditions and [writing a]...test case} (P356). Others opted to \emph{"use words (tags) rather than sentences"} (P206).

\participantQuote{Be incredibly specific with the instructions and write them as precisely as I would for a stupid collaborator.}{170}

\paragraph{\textbf{No strategy} (44$\times$)}
Many participants reported not employing any strategy, as they found \tool{}s to provide helpful suggestions without needing to perform specific actions.

\participantQuote{Nothing, I just review the suggestions as they come up.}{268}

\paragraph{\textbf{Adding code} (36$\times$)}
Participants often reported consciously writing additional code as context for the \tool\xspace to later complete. Participants did this to \emph{"make some context"} (P117) and provide a \emph{"hint to [improve] the code generation"} (P93).

\participantQuote{Write a partial fragment of the code I think is...correct.}{166}

\paragraph{\textbf{Following conventions} (24$\times$)}
Many participants also resorted to following common conventions, such as \emph{"communities' rules and design patterns"} (P157), \emph{"well-named variables"} (P366), or \emph{"[giving] the function a very precise name"} (P254). Participants even viewed the generated code as a source of code with proper conventions:

\participantQuote{Proper naming conventions also helps... Since these tools learn from excellent code, I should also write code that follows conventions, this can make tools easily find the right result.}{224}

\paragraph{\textbf{Breaking down instructions} (18$\times$)}
Participants also reported breaking down the code logic or prompts into shorter, more concise statements by explaining the functionality step-by-step. Examples include \emph{"break[ing] the problem into smaller parts"} (P166) and \emph{"split[ting] the sentence to be shorter"} (P167).

\participantQuote{You have to break down what you're trying to do and write it in steps, it can't do too much at once.}{126}

\paragraph{\textbf{Existing code context} (18$\times$)}
Participants developed mental models of these tools~\cite{cheng2022would}, as they reported leveraging existing code as additional data for the \tool{} to use, such as by \emph{"opening files for context"} (P274). Participants reported specifically using \tool{}s only when there was sufficient existing code context:

\participantQuote{I try to use it at advanced stages of my project, where it can give better suggestions based on my project's history.}{111}

\paragraph{\textbf{Prompt engineering} (13$\times$)}
Some participants iteratively changed their inputs to query the tool.  such as \emph{"changing the prompt/comment to simpler sentences"} (P82) or \emph{"tweak[ing] the comments...to [be more] interactive...for the specific task"} (P80).

\participantQuote{If the code generated does not satisfy me, I will edit the comments.}{150}

\mybox{\faArrowCircleRight\xspace\textbf{Key findings:} 
Participants who were GitHub Copilot users reported a median of 30.5\% of their code being written with its help (\#1). 
The most important reasons for using \tool{}s were for autocomplete, completing programming tasks faster, or skipping going online to recall syntax (\#2). 
Participants successfully used these tools to generate code that was repetitive or had simple logic.
Participants reported the most important reasons for not using \tool{}s were because the code that the tools generated did not meet functional or non-functional requirements and because it was difficult to control the tool (\#3).
}

\begin{table*}
  \centering

\small

\caption{How frequently participants report usability issues occurring while using \tool{}s.}
\label{tab:usability-issues}
\begin{tabular}{p{0.02\linewidth}p{0.47\linewidth}|p{0.45\linewidth}}
\toprule
& \textbf{Situation} & \textbf{Distribution} \\
\midrule
& \textbf{\emph{A. Usability issues}} & \\
S1 & \hangindent=1em I don't know what part of my code or comments the code generation tool is using to make suggestions.  & \frequencybarchart{0.08}{0.22}{0.22}{0.25}{0.23}{0}{30\%}{48\%}\\
S2 & \hangindent=1em I give up on incorporating the code created by a code generation tool and write the code myself. &  \frequencybarchart{0.07}{0.21}{0.37}{0.29}{0.06}{0}{28\%}{35\%} \\
S3 & I have trouble controlling the tool to generate code that I find useful. & \frequencybarchart{0.06}{0.2}{0.26}{0.29}{0.19}{0}{26\%}{48\%}\\
S4 & I find the code generation tool's suggestions too distracting. & \frequencybarchart{0.05}{0.18}{0.34}{0.31}{0.13}{0}{23\%}{44\%}\\
S5 & I have trouble evaluating the correctness of the generated code. & \frequencybarchart{0.05}{0.18}{0.25}{0.39}{0.13}{0}{23\%}{52\%} \\
S6 & \hangindent=1em I have difficulty expressing my intent or requirements through natural language to the tool. &  \frequencybarchart{0.05}{0.17}{0.42}{0.26}{0.1}{0}{22\%}{36\%}\\
S7 & I find it hard to debug or fix errors in the code from code generation tools. & \frequencybarchart{0.05}{0.12}{0.22}{0.38}{0.23}{0}{17\%}{61\%} \\
S8 & I rely on code generation tools too much to write code for me. &  \frequencybarchart{0.04}{0.11}{0.24}{0.27}{0.35}{0}{15\%}{67\%}\\
S9 & I have trouble understanding the code created by a code generation tool. & \frequencybarchart{0.01}{0.05}{0.22}{0.42}{0.3}{0}{5.6\%}{45\%}\\
\midrule
& \textbf{\emph{B. Reasons for not understanding code output}} & \\
S10 & The generated code uses APIs or methods I don't know. & \frequencybarchart{0.03}{0.22}{0.42}{0.39}{0.04}{0}{25\%}{43\%} \\
S11 & The generated code is too long to read quickly. & \frequencybarchart{0.05}{0.18}{0.33}{0.39}{0.06}{0}{23\%}{45\%} \\
S12 & \hangindent=1em The generated code contains too many control structures (e.g., loops, if-else statements). & \frequencybarchart{0.04}{0.15}{0.4}{0.4}{0.04}{0}{19\%}{44\%} \\
\midrule
& \textbf{\emph{C. Methods of evaluating code output}} & \\
S13 & \hangindent=1em Quickly checking the generated code for specific keywords or logic structures &  \frequencybarchart{0.34}{0.4}{0.16}{0.08}{0.02}{0}{74\%}{10\%} \\
S14 & Compilers, type checkers, in-IDE syntax checkers, or linters &  \frequencybarchart{0.42}{0.29}{0.16}{0.1}{0.04}{0}{71\%}{14\%} \\
S15 & Executing the generated code &  \frequencybarchart{0.36}{0.33}{0.17}{0.11}{0.03}{0}{69\%}{14\%} \\
S16 & Examining details of the generated code's logic in depth &  \frequencybarchart{0.28}{0.36}{0.22}{0.11}{0.04}{0}{64\%}{15\%} \\
S17 & Consulting API documentation &  \frequencybarchart{0.13}{0.25}{0.34}{0.18}{0.1}{0}{38\%}{28\%} \\
\midrule
& \textbf{\emph{D. Methods of modifying code output}} & \\
S18 & \hangindent=1em When a code generation tool outputs something I don't want, I'm able to modify it to something I want. &  \frequencybarchart{0.24}{0.39}{0.26}{0.1}{0.02}{0}{63\%}{12\%} \\
S19 & \hangindent=1em I successfully incorporate the code created by a code generation tool by changing the generated code. &  \frequencybarchart{0.12}{0.5}{0.29}{0.07}{0.02}{0}{62\%}{9.1\%} \\
S20 & I use the code created by a code generation tool as-is. & \frequencybarchart{0.04}{0.41}{0.32}{0.2}{0.04}{0}{44\%}{24\%} \\
S21 & \hangindent=1em I successfully incorporate the code created by a code generation tool by changing the code or comments around it and regenerating a new suggestion. &  \frequencybarchart{0.1}{0.3}{0.31}{0.22}{0.08}{0}{40\%}{30\%} \\
\midrule
& \textbf{\emph{E. Reasons for giving up on code output}} & \\
S22 & The generated code doesn't perform the action I want it to do. &  \frequencybarchart{0.1}{0.33}{0.35}{0.19}{0.03}{0}{43\%}{22\%} \\
S23 & \hangindent=1em The generated code doesn't meet functional or non-functional (e.g., security, performance) requirements that I need. &  \frequencybarchart{0.04}{0.3}{0.39}{0.23}{0.05}{0}{34\%}{28\%} \\
S24 & The generated code's style doesn't match my project's. &  \frequencybarchart{0.06}{0.16}{0.31}{0.34}{0.14}{0}{22\%}{48\%} \\
S25 & The generated code contains too many defects. &  \frequencybarchart{0.03}{0.18}{0.38}{0.36}{0.06}{0}{21\%}{42\%} \\
S26 & The generated code uses an API I know, but don't want to use. &  \frequencybarchart{0.03}{0.14}{0.28}{0.37}{0.18}{0}{17\%}{55\%} \\
S27 & I don't understand the generated code well enough to use it. &  \frequencybarchart{0.02}{0.1}{0.19}{0.47}{0.22}{0}{12\%}{69\%} \\
S28 & The generated code is too complicated. &  \frequencybarchart{0.01}{0.09}{0.22}{0.46}{0.22}{0}{10\%}{68\%} \\
S29 & The generated code uses an API I don't know. &  \frequencybarchart{0.03}{0.07}{0.3}{0.37}{0.23}{0}{10\%}{59\%} \\
\midrule
\multicolumn{3}{c}{\mylegend{Always}{green3} \mylegend{Often}{green2} \mylegend{Sometimes}{gray1}\mylegend{Rarely}{red2} \mylegend{Never}{red3}} \\
\bottomrule
\end{tabular}
\end{table*}
\section{Usability of \tool{}s}
\label{sec:usability-challenges}
In this section, we present our findings on what challenges developers encounter while interacting with \tool{}s. We first report the frequency of usability issues (Section~\ref{sec:usability-issues}). To better understand these challenges, we explore the practices of users in understanding (Section~\ref{sec:understanding-code}), evaluating (Section~\ref{sec:evaluating-code}), modifying (Section~\ref{sec:modifying-code}), and giving up (Section~\ref{sec:giving-up-code}) on outputted code.

\begin{table*}
  \centering

\small

\caption{Participants' level of concern on issues related to \tool{}s.}
\label{tab:concerns}
\begin{tabular}{p{0.02\linewidth}p{0.47\linewidth}|p{0.45\linewidth}}
\toprule
& \textbf{Concern} & \textbf{Distribution} \\
\midrule
C1 & \hangindent=1em Code generation tools produce code that infringe on intellectual property. & \importancebarchart{0.25}{0.21}{0.22}{0.16}{0.16}{0}{46\%}{32\%}\\
C2 & Code generation tools have access to my code. & \importancebarchart{0.2}{0.21}{0.21}{0.17}{0.21}{0}{41\%}{38\%}\\
C3 & Code generation tools do not generate proprietary APIs or code. & \importancebarchart{0.14}{0.15}{0.26}{0.18}{0.28}{0}{29\%}{46\%}\\
C4 & Code generation tools may produce open-source code. & \importancebarchart{0.12}{0.17}{0.18}{0.14}{0.39}{0}{29\%}{53\%}\\
\midrule
\multicolumn{3}{c}{\mylegend{Very concerned}{blue2} \mylegend{Concerned}{blue1} \mylegend{Moderately concerned}{gray1}\mylegend{Slightly concerned}{orange1} \mylegend{Not concerned at all}{orange2}} \\
\bottomrule
\end{tabular}
\end{table*}

\subsection{Usability issues}
\label{sec:usability-issues}
We asked participants to rate how frequently certain usability issues occurred while they used \tool{}s (see Table~\ref{tab:usability-issues}-A). The biggest challenges participants reported facing were not knowing what part of the input influenced the output (\situation{1}), giving up on using outputted code (\situation{2}), and having trouble controlling the model (\situation{3}), as 30\%, 28\%, and 26\% of participants encountered these situations often. Meanwhile, participants had the least trouble with understanding the code generated by the tool (\situation{9})--only 5.6\% of participants frequently encountered this issue, despite it being discussed in prior literature~\cite{vaithilingam2022expectation}. 

\subsection{Understanding outputted code}
\label{sec:understanding-code}
We asked participants who reported having trouble understanding the outputted code to rate the reasons why (see Table~\ref{tab:usability-issues}-B). 25\% of participants said it was often because the outputted code used unfamiliar APIs (\situation{10}). Meanwhile, 23\% and 19\% of participants stated it was often due to the code being too long to read quickly (\situation{11}) and the code having too many control structures (\situation{12}) respectively.

\subsection{Evaluating outputted code}
\label{sec:evaluating-code}
We asked participants how they evaluated generated code (see Table~\ref{tab:usability-issues}-C). The order of the evaluation methods by frequency closely related to how time-consuming each method was reported to be. Participants often reported using quick visual inspections of the code (\situation{13}, 74\%), static analysis tools like syntax checkers (\situation{14}, 71\%), executing the code (\situation{15}, 69\%), and examining the details of the outputted code's logic in depth (\situation{16}, 64\%). However, participants reported frequently consulting API documentation at a lower rate (\situation{17}, 38\%).

\subsection{Modifying outputted code}
\label{sec:modifying-code}
We asked participants how they modified the generated code (see Table~\ref{tab:usability-issues}-D). Participants overall reported regularly having success with modifying the outputted code (\situation{18}, 63\%), most often by changing the generated code itself (\situation{19}, 62\%) rather than by changing the input context (\situation{20}, 40\%). Additionally, a smaller proportion of participants (\situation{21}, 44\%) often used the generated code as-is.

\subsection{Giving up on outputted code}
\label{sec:giving-up-code}
We asked participants who reported giving up on outputted code to rate the reasons why (see Table~\ref{tab:usability-issues}-E). The two major reasons were that the generated code did not perform the intended action (\situation{22}) and because the code did not meet functional or non-functional requirements (\situation{23})--43\% and 34\% of participants frequently encountered these situations respectively. 
The least salient reasons why participants gave up on using generated code was that they did not understand the outputted code (\situation{27}), that they found the output too complicated (\situation{28}), and that the outputted code used unfamiliar APIs (\situation{29}). This was regularly encountered by 12\%, 10\%, and 10\% of participants respectively.

\mybox{\faArrowCircleRight\xspace\textbf{Key findings:} 
The most frequent usability challenges participants reported encountering were understanding what part of the input caused the outputted code, giving up on using the outputted code, and controlling the tool's generations (\#4). 
Participants most often gave up on outputted code because the code did not perform the intended action or did not account for certain functional and non-functional requirements (\#5).}

\section{Additional Feedback}
\label{sec:user-feedback}
We present our results on what additional feedback developers have to improve their experiences with \tool{}s. We discuss general concerns that participants had about these tools (Section~\ref{sec:general-concerns}) and participants' responses on how they would improve them (Section~\ref{sec:improving-tools}).

\subsection{General concerns}
\label{sec:general-concerns}
We asked all participants to rate their level of concern on issues related to \tool{}s (see Table~\ref{tab:concerns}), which were derived from ~\citet{cheng2022would} and our survey pilots. 
Participants overall seemed most concerned about their own and others' intellectual property--they most frequently described feeling concerned over \tool{}s producing code that infringed on intellectual property (\concern{1}, 46\%) and the tools having access to their code (\concern{2}, 41\%). 
In contrast, participants seemed less worried about concerns more specific to working in commercial contexts; 29\% of participants reported feeling concerned about \tool{}s not generating proprietary APIs (\concern{3}) as well as generating outputted code that contained open-source code (\concern{4}).

\subsection{Improving \tool{}s}
\label{sec:improving-tools}
We asked participants to describe feedback they would provide to \tool{}s to make their output better. We identified 8 types of feedback, which we elaborate on below.

\paragraph{\textbf{User feedback} (52$\times$)}
Most frequently, participants wanted to provide feedback to the \tool{} for it to learn from. Some wanted to correct the outputted code as feedback, while others wanted to teach the model their personal coding style. While some participants wanted to directly provide feedback in natural language, others preferred code: \emph{ "Maybe...code [of] my correct answer. I don't...want to explain in natural language."} (P201). Meanwhile, others suggested rating the output with \emph{"like/dislike buttons...to not get distracted from actual work"} (P52).

\participantQuote{Automatic feedback based on code correction made by the developer.}{57}

\participantQuote{Maybe more personaliz[ation]...I have my own code style, so I will need...time to modify the code into my style.}{102}

\paragraph{\textbf{Better understanding of code context} (20$\times$)}
Participants also reported wanting \tool{}s to have additional understanding of code context, such as learning from \emph{"context from other files on the same workspace"} (P12). Others wanted these tools to have a deeper understanding of certain nuances behind APIs and programming languages, such as when \emph{"the code is using [a] deprecated API"} (P88).

\participantQuote{To be able to better describe the contexts of our projects during creation. For a better understanding of our code generator.}{208}

\paragraph{\textbf{Tool configuration} (17$\times$)}
A few participants wanted to change the tool's settings. This included \emph{"distinguish[ing when to do] long code generation and short code [generation]"} (P240), having \emph{"adjustable parameters"} (P177), or reducing the frequency of suggestions. This could assist the model in adapting to whether the developer was in \emph{acceleration mode}--associated with short completions--or \emph{exploration mode}--associated with long completions~\cite{barke2022grounded}.

\participantQuote{I'd like to be able to ask it to calm down sometimes instead of constantly trying to suggest random stuff.}{122}

\paragraph{\textbf{Natural language interactions} (16$\times$)}
Some participants wanted opportunities for interaction via natural language. Inspired by ChatGPT~\cite{openai2023chatgpt}, several participants mentioned chat-based interactions: \emph{"would be nice if we could give feedback to it like how we chat with chatGPT"} (P39). 

\participantQuote{To comment on the resulting code the tool generates, and let the tool reiterate from such previously generated result, but with my comments.}{166}

\paragraph{\textbf{Code analysis} (13$\times$)}
As discussed in prior work~\cite{barke2022grounded}, some participants also wanted further analysis on the generated code for functional and syntactic correctness, as \emph{"[making] any basic grammatical mistakes or spelling mistakes...would be considered unreliable"} (P105). 

\participantQuote{Add extra checks to outputted code to ensure it resembles the input given and that the outputted code is complete and can be run. Often the outputted code that I am given is incomplete, lacks the ability to run or [be] tested immediately.}{158}
        
\paragraph{\textbf{Explanations} (11$\times$)}
Some participants wanted explanations for additional context of the generated code, such as \emph{"sourcing...the suggestions"} (P58) or \emph{"link[ing] direct[ly] to documentation"} (156).

\participantQuote{These tools must show where the code snippet comes from and include the code link of snippet, license, author name if available for better references for that specific code.}{281}

\paragraph{\textbf{More suggestions} (9$\times$)}
Consistent with prior work~\cite{barke2022grounded}, a few participants wanted to have the model regenerate or provide more than one suggestion, such as by having the \emph{"possibility to shuffle between code snippets"} (P177).

\participantQuote{Maybe multiple suggestions and then I pick the best.}{149}

\paragraph{\textbf{Accounting for non-functional requirements} (8$\times$)}
Some participants requested \tool{}s to generate code that addressed non-functional requirements, such as \emph{"time complexity"} (P191). Other participants wanted more readable code:

\participantQuote{Sometimes AI suggest code [with] one lines or short hand logic, which is difficult to read and understand.}{98}

\mybox{\faArrowCircleRight\xspace\textbf{Key findings:} Participants were most concerned about potentially infringing on intellectual property and having a tool have access to their code.
Participants reported wanting to improve \tool{}s' output by having users directly provide feedback to correct or personalize the tool or by teaching the underlying model to have a better understanding of code context (\#6).
They also wanted more opportunities for natural language interaction with these tools.
}

\section{Threats to Validity}
\label{sec:threats-to-validity}
\subsubsection*{Internal validity}
Memory bias may influence the internal validity of the study, as the survey questions required participants to recall their experiences with \tool{}s. We addressed this threat by asking participants to consider their experiences with these tools with respect to a specific project in order to ground participant responses with a concrete experience.

Study participants may also misunderstand the wording of some of the survey questions. To reduce this threat, we piloted the survey 11 times with developers with a focus on the clarity of the survey questions and updated the survey based on their feedback.

\subsubsection*{External validity}
Any empirical study may have difficulties in generalizing~\cite{flyvbjerg2006five}. To address this, we sample from a set of participants who are diverse in terms of geographic location and software engineering experience. 
However, our study may still struggle with sampling bias. This is because we sampled from GitHub projects that were related to \tool{}s, such as GitHub Copilot and Tabnine. 
Thus, our sample \added{largely represents people who are enthusiastic about these tools. Further, our sample} does not specifically sample individuals who are not interested in \tool{}s, so this population may be underrepresented within our study. 
\added{Therefore, our sample may not be representative of all users of \tool{}s.}

Because the survey was deployed in January 2023, participants provided responses based on their experiences with \tool{}s at the time. Thus, some aspects may not be relevant to future versions of these tools that perform differently.

\subsubsection*{Construct validity}
Many survey questions asked participants to provide subjective estimates of the frequency of encountering certain situations or using specific tools. Thus, these estimates may not be accurate. Collecting \emph{in-situ} data in future studies, such as in ~\cite{mozannar2022reading} and ~\cite{ziegler2022productivity}, would be more appropriate to evaluate the frequency of these events. We report measurements on perceived frequency as a proxy for the importance of each usability challenge--following best practices in human factors in software engineering research~\cite{myers2016programmers}--rather than the ground truth on the usability challenge's frequency.

\added{
\subsubsection*{Ethical Considerations}
An important component of this research study was gathering a sufficiently large number of responses to our survey. Our goal was to receive 385 survey responses, so that we could achieve a 95\% confidence level with a 5\% margin of error with our sample. 

Given our recruitment method needed to result in a large number of responses from programmers, traditional methods of recruitment used in smaller-scale user studies were not practical for our study. Snowball sampling was unlikely to yield the scale of responses that were necessary, while recruiting student programmers from our institution or using traditional crowd-sourcing platforms (e.g., Amazon Mechanical Turk) would not target a representative population of developers. Therefore, we followed prior research in the past 10 years published in top software engineering conferences (\cite[e.g.,][]{huang2021leaving, liang2022understanding, gousios2016work, gousios2015work}) that utilized large-scale participant recruitment from populations on GitHub that achieved a sufficient number of survey responses. However, community standards following this recruitment method have recently shifted. Recent work from  \citet{tahaei2022lessons} has noted limitations in this method, as mining emails from GitHub is not encouraged by the platform. We advise future work to not use our recruitment strategy and instead follow \citet{tahaei2022lessons}'s recommendation in using the crowdsourcing platform, Prolific~\cite{prolific2023prolific}, as it is a more sustainable way of gathering survey responses from developers at scale.
}

\section{Discussion \& Future Work}
\label{sec:discussion-future-work}
The findings from our study overlap with prior usability studies of \tool{}s~\cite[e.g.,][]{ziegler2022productivity, barke2022grounded, vaithilingam2022expectation,bird2022taking}. In this section, we discuss these works in relation to our results. This produces several implications for future work, which we elaborate on further. 

\subsection{Implications}

\subsubsection*{Acceleration mode versus exploration mode}
\citet{barke2022grounded} found that users of \tool{}s, such as GitHub Copilot, use the tools in two main modes: \emph{acceleration mode}, where the developer knows what code they would like to write and uses the tool to complete the code more quickly, or \emph{exploration mode}, where the developer is unsure of what to write and would like to visit potential options. Our results support this theory of \tool{} usage, as both \emph{acceleration mode} and \emph{exploration mode} emerge as themes in our results. In particular, these modes appear when developers use \tool{}s (e.g., \qcode{repetitive code}, \qcode{code with simple logic}, \qcode{autocomplete}, \qcode{recalling} versus \qcode{proof-of-concepts}), why developers use these tools (e.g., autocompleting (\motivation{1}), finishing programming tasks faster (\motivation{2}), not needing to go online to find code snippets (\motivation{3}) versus discovering potential ways to write a solution (\motivation{4}), finding an edge case (\motivation{5})), and how developers interacted with the tool to produce better suggestions (e.g., \qcode{no strategy}, \qcode{following conventions}, \qcode{adding code} versus \qcode{clear explanations}).  

We further augment ~\citet{barke2022grounded}'s theory by finding that aspects related to \emph{acceleration mode} are represented within our data more than aspects related to \emph{exploration mode}. For example, \qcode{repetitive code} (78$\times$), \qcode{code with simple logic} (68$\times$), and \qcode{autocomplete} (28$\times$), all occur more frequently than \qcode{proof-of-concepts} (20$\times$) as situations when participants successfully used \tool{}s. Additionally, participants rated \motivation{1} (86\%), \motivation{2} (76\%), and \motivation{3} (68\%) to be important reasons for using \tool{}s at higher rates than \motivation{4} (50\%) and \motivation{5} (36\%). This suggests that developers may value \emph{acceleration mode} over \emph{exploration mode}.

\subsubsection*{Chatbots as \tool{}s}
Our results also indicate a potential for \tool{} users to rely more on chat-based interactions, following the recent rise of powerful chatbots such as ChatGPT~\cite{openai2023chatgpt}. 
6\% of our participants explicitly wrote that they used ChatGPT as an \tool{}, and a popular feedback was to provide more opportunities for \qcode{natural language interactions}. 
While recent work shows promise in this method of interaction with \tool{}s~\cite{robe2022pair, ross2023programmer}, it also raises additional questions of when these interaction methods should be applied. 
Understanding \emph{when} developers should rely on these interactions is fundamentally a usability question that cannot be addressed through technological advances alone, as it is unclear how to balance this interaction mode with users' cognitive load.
While participants seemed to prefer \emph{acceleration mode} over \emph{exploration mode}, our results also indicate that some users may be amenable to using chat;
this is because providing \qcode{clear explanations}, often in natural language, was the most cited strategy to having \tool{}s produce the best output.

\subsubsection*{Developers using \tool{}s to learn APIs and programming languages}
The findings from our study indicate the potential for developers using \tool{}s to learn APIs and programming languages. 
Learning is a fundamental action in software engineering~\cite{ford2017characterizing} and is independent of any technological innovation. Further, it is an important skill for developers~\cite{liang2022understanding,li2015makes,li2020distinguishes,baltes2018towards}.
While developers previously used online resources, such as documentation~\cite{rao2020analyzing}, StackOverflow~\cite{herbsleb2001global,mamykina2011design}, or blogs~\cite{storey2014r} to learn how to use new technologies, our study participants often favored \tool{}s over these resources for both \qcode{recalling} and \qcode{learning} syntax of APIs and programming languages. 

\subsubsection*{Aligning \tool{}s to developers}
Our results indicate that there are several opportunities in aligning \tool{}s to the needs of developers. 
Giving up on incorporating code (\situation{2}) was the most common usability issue encountered and it often occurred because the code did not perform the correct action (\situation{22}). Future work could mitigate this issue by designing new metrics (e.g., ~\cite{dibia2022aligning}) to increase developer-tool alignment.

Further, one emergent theme to align these tools with developers is by giving developers more control over the tools' outputs.
In our study, the most frequent usability issues encountered were not knowing why code was outputted (\situation{1}, 30\%) and having trouble controlling the tool (\situation{3}, 26\%). Participants also often reported not using these tools due to difficulties controlling the tool (\motivation{7}, 48\%). Additionally, the most frequent feedback provided was accepting \qcode{user feedback} to correct the tool. 
Thus, future work should investigate techniques to allow users to better control \tool{}s, such as through interactive machine learning approaches~\cite{amershi2014power}.

Another theme that emerged was the need for \tool{}s to account for non-functional requirements in the generation. 
It was mentioned within the feedback that study participants had for the tools (\qcode{accounting for non-functional requirements}) and was a reason why participants did not use them (\motivation{6}, 54\%) or gave up on generated code (\situation{23}, 34\%). 
Therefore, future work should investigate avenues for incorporating non-functional requirements--such as readability and performance--into the generation, which could help increase developers' adoption of these tools. One such example is GitHub's recent project, Code Brushes~\cite{github2023codebrushes}.

\added{
\subsection{Takeaways}
These implications affect both software engineering researchers and practitioners. Below we describe how our findings apply to these populations and discuss opportunities for future work.

\subsubsection*{For practitioners \& tool users}
Our findings point to strategies for practitioners to use \tool{}s more efficiently, which could potentially boost productivity. For instance, software practitioners could make additional efforts to provide \qcode{clear explanations} to prompt the \tool{} effectively. Practitioners could consider combining this with \qcode{adding code} or \qcode{following conventions} (e.g., programming conventions) to get the highest quality output possible.

Additionally, our results reveal new use cases of \tool{}s for practitioners. Rather than using these tools for only autocompletion, software practitioners could consider using them for \qcode{quality assurance} (e.g., generating test cases) as well as learning new APIs or programming languages.

\subsubsection*{For researchers \& tool creators}
The results from our study reveal several interesting directions for future research, which could be incorporated into \tool{}s. For example, given participants' reliance on ChatGPT and \qcode{natural language interactions}, future work could investigate methods for supporting chat-based interactions without impacting developers' efficiency and flow while programming~\cite{forsgren2021space}. Additionally, future work could investigate how developers learn new technologies with \tool{}s and design experiences that help support developer \qcode{learning}.

Another line of research is to study how to improve \tool{}s' alignment with developers. This is unlikely to be resolved entirely through modeling improvements, as human developers must be able to articulate requirements and evaluate solutions for any given problem. 
However, this is challenging, as software design and implementation are notoriously complex. Software solutions and problems can co-evolve with one another~\cite{van2016decision}, and software design knowledge can be implicit~\cite{liang2023qualitative}. Thus, facilitating ways for developers to explicitly describe their software design knowledge to these tools is a challenge to address.

Finally, future work should also investigate new interaction techniques to support \emph{acceleration mode} specifically, given participants' emphasis on this type of usage of \tool{}s. Following design recommendations for generative AI in creative writing contexts~\cite{clark2018creative}, these interaction techniques should require minimal cognitive effort for developers to prevent distracting them from their tasks. Study participants described favoring implicit interactions with \tool{}s over explicit ones:

\participantQuote{Automatic feedback. The tool knows whether I choose...to apply its suggestions. Because it won't distract me.}{246}

\participantQuote{Feedback...is important, but I'm not sure I want to invest time in "teaching" the tool.}{111}
}

\section{Conclusion}
In this study, we investigated the usability of \tool{}s, such as GitHub Copilot. We performed an exploratory qualitative study by surveying 410 developers on their usage of \tool{}s to better understand their usage practices and uncover important usability challenges they encountered. 

We find that developers are most motivated to use \tool{}s because of the tools' ability to autocomplete, help finish programming tasks quickly, and recall syntax, rather than helping developers brainstorm potential solutions for problems they are facing.
We also find that while state-of-the-art \tool{}s are highly performant, there is a gap between developers' needs and the tools' output, such as accounting for non-functional requirements in the generation.

Our findings indicate several potential directions for \tool{}s, such as designing interaction techniques that provide developers with more control over the tool's output. To facilitate replication of this study, the survey instrument and codebooks are included in the supplemental materials for this work~\cite{supplemental-materials}.

\begin{acks}
We thank our survey participants for their wonderful insights. We also thank Alex Cabrera, Samuel Estep, Vincent Hellendoorn, Kush Jain, Christopher Kang, Millicent Li, Christina Ma, Manisha Mukherjee, Soham Pardeshi, Daniel Ramos, Sam Rieg, and others for their feedback on the study. We also give a special thanks to Mei \meiicon, an outstanding canine software engineering researcher, for providing support and motivation throughout this study. Jenny T. Liang was supported by the National Science Foundation under grants DGE1745016 and DGE2140739. Brad A. Myers was partially supported by NSF grant IIS-1856641. Any opinions, findings, conclusions, or recommendations expressed in this material are those of the authors and do not necessarily reflect the views of the sponsors.
\end{acks}

\bibliographystyle{ACM-Reference-Format}
\bibliography{acmart}

\end{document}